\documentclass[pre,eqsecnum,showpacs,byrevtex,amsmath,amssymb,nofootinbib]{revtex4}
\usepackage{amsmath,graphicx}

\begin{document}

\title{Collapse or Swelling Dynamics of Homopolymer Rings: \\
  Self-consistent Hartree approach}
 
\author{Vakhtang G. Rostiashvili$^{1,2}$, Nam-Kyung Lee$^{1}$ and Thomas A.
  Vilgis$^{1,2}$}\email{vilgis@mpip-mainz.mpg.de}
\affiliation{$^1$ Max Planck Institute for Polymer Research\\
  10 Ackermannweg, 55128 Mainz, Germany.\\
  $^2$ Laboratoire Europ\'een Associ\'e, Institute Charles Sadron\\
  6 rue Boussingault, 67083 Strasbourg Cedex, France.}

\begin{abstract}
  We investigate by the use of the Martin - Siggia - Rose generating functional
  technique and the self - consistent Hartree approximation, the dynamics of
  the ring homopolymer collapse (swelling) following an instantaneous change into a
  poor (good) solvent conditions.The equation of motion for the time dependent
  monomer - to - monomer correlation function is systematically derived. It is
  argued that for describing of the coarse - graining process (which neglects
  the capillary instability and the coalescence of ``pearls'') the Rouse mode
  representation is very helpful, so that the resulting equations of motion
  can be simply solved numerically. In the case of the collapse this solution
  is analyzed in the framework of the hierarchically crumpled fractal picture,
  with crumples of successively growing scale along the chain. The presented
  numerical results  are in line with the corresponding simple scaling argumentation
  which in particular shows that the characteristic collapse time of a segment
  of length $g$ scales as $t^* \sim \zeta_0 g/\tau$ (where $\zeta_0$ is a bare
  friction coefficient and $\tau$ is a depth of quench). In contrast to the
  collapse the globule swelling can be seen (in the case that topological
  effects are neglected) as a homogeneous expansion of the globule interior.
  The swelling of each Rouse mode as well as gyration radius $R_g$ is
  discussed.
\end{abstract}
\pacs{61.25.Hq,82.35.Lr,36.20.-r} 
\maketitle

\section{Introduction}
The equilibrium theory of the coil - globule transition is one of the
major issues in polymer physics. Many theories have been developed to
understand this physical phenomena in more detail.
\cite{Gennes,Gennes1,Grosberg,Kremer}.  The theory for kinetics of
this transition
\cite{abrams01,Gennes2,Pitard,Pitard1,Pitard2,Brochard,Halperin,Klushin,Byrne,Timoshenko,Shakh,Chang}
as well as the pertinent experiment \cite{Wu}
is the subject which has recently attracted a broad interest.
One of the main motivations for these studies is that the first stage of a
protein folding process is believed to be a fast collapse to a compact but
nonspecific state. This has been shown for example by lattice Monte Carlo
simulations \cite{Socci}.  The first stage of folding appears as sequence -
independent and therefore the process is similar to the collapse of a
homopolymer. Generally the collapse of a polymer chain can be addressed to an
(abrupt) change of the second virial coefficient from the good solvent regime
$v>0$ to the poor solvent regime $v<0$. The second virial coefficient depends 
on the temperature and has a Boyle point $v=0$ at the so called $\theta$
temperature. The resulting attractive two body interactions requires at least
the third virial (always positive)   which prevents the chain from
the collapse to a single point.

In de Gennes' ``expanding sausage model'' \cite{Gennes2} a flexible chain
changes its conformations on a shortest scale through the formation of
crumples after being quenched to poor solvent conditions. Then crumples are
formed on a larger scale, resulting in a sausage - like shape, which
eventually leads to a compact globule. One can see  that this model is
translational invariant along the chain backbone, provided that the chain has
cyclic boundary conditions. This simplifies the problem and allows us later on
the use of simple Rouse transformations \cite{Pitard,Pitard1,Doi} of the chain
coordinates.

Concerning de Gennes' model, it is shown by Brownian dynamics simulations
\cite{Byrne}, that the ``sausages'' becomes unstable with respect to capillary
waves. Consequently a so called pearl necklace structure is formed. This
brought about a number of publications where scaling arguments
\cite{Brochard,Halperin,Klushin} and the Gaussian self - consistent approach
\cite{Timoshenko} have been put forward. Actually the pearl necklace formation
breaks the translational invariance along the chain backbone and complicates
these issues.

Alternative recent phenomenological considerations are put forward by joining
scaling argument \cite{abrams01} and computer simulations
\cite{abrams01,Chang}. It was argued \cite{abrams01} that the relaxations
times related with the capillary instability as well as with coalescence of
pearls are short compared to the characteristic time on which the ``sausage''
change its configuration. Since the positions of pearls along the
chain are  random they  can be  averaged  over. The resulting ``sausage'' can be seen as an envelop (which
has the form of a flexible cylinder) of the pearl necklaces and the overall
chain configurations are composed by random walks of sausages.  This picture
successfully explains the so called coarse - graining process shortly after
the quench to a range of temperature below $\theta$ - point, but above
eventual glassy globular relaxation regimes \cite{Pitard2,Dokholyan,Rost}.
This argumentation reconciles in a sense the two scenarios mentioned above,
but the microscopic picture is still lacking.
Therefore instead of considering the processes of formation and coalescence of
pearls we concentrate here on the coarse - grained dynamics of the
``sausage''. This model, as it was already mentioned,  is homogeneous in the sense that the translational
invariance  along the chain backbone is effectively assured during the process of the collapse.

In this paper we provide the microscopic theory for the coil to globule
transition. In order to get insight and some intuitive picture for the 
later discussion we start from the scaling consideration. Then we
study the Langevin dynamics of the problem based on  the Martin - Siggia - Rose (MSR) generating functional
technique together with the self - consistent Hartree approximation
\cite{Horner,Shapir,Rost1}. It is quite important for the judgment of the
final results how the low molecular solvent dynamics is treated. In this paper
we restrict our consideration to the random phase approximation (RPA) which is
well - known in the context of the theory of both polymeric \cite{Vilgis} and low -
molecular  \cite{Boon} systems.  The RPA fails to account the
hydrodynamic interaction because in the hydrodynamic regime
collisions dominate and keep the solvent in a state of local
thermodynamic equilibrium (see e.g. Sec.6.5 in ref. \cite{Boon}), so
that we leave this subject for the future publications. As a main result the
generalized dynamical equation for the collapsed (or swelled) chain has been
derived. The relaxation laws for the early and late stages are investigated
analytically, whereas the whole numerical solution is also done and thoroughly
discussed. 

\section{Scaling}

\subsection{Collapse}

Let us first consider the dynamical time scales for the ring  polymer collapse under
Rouse dynamics conditions. By this assumption we neglect certain physical
properties, such as capillary instability and long ranged hydrodynamic
interactions.  In this first section, our main goal is to provide a
corresponding scaling picture for a Rouse chain which we are going to discuss
with more refined analytical methods below. Nevertheless the principal times
scales are fixed.  The relaxation of the each length scale is clearly
associated with the relaxation of mode as we will show later.

\begin{figure}[h]
  \begin{center}
  \begin{minipage}{10cm}
    \centerline{\includegraphics[width=10cm]{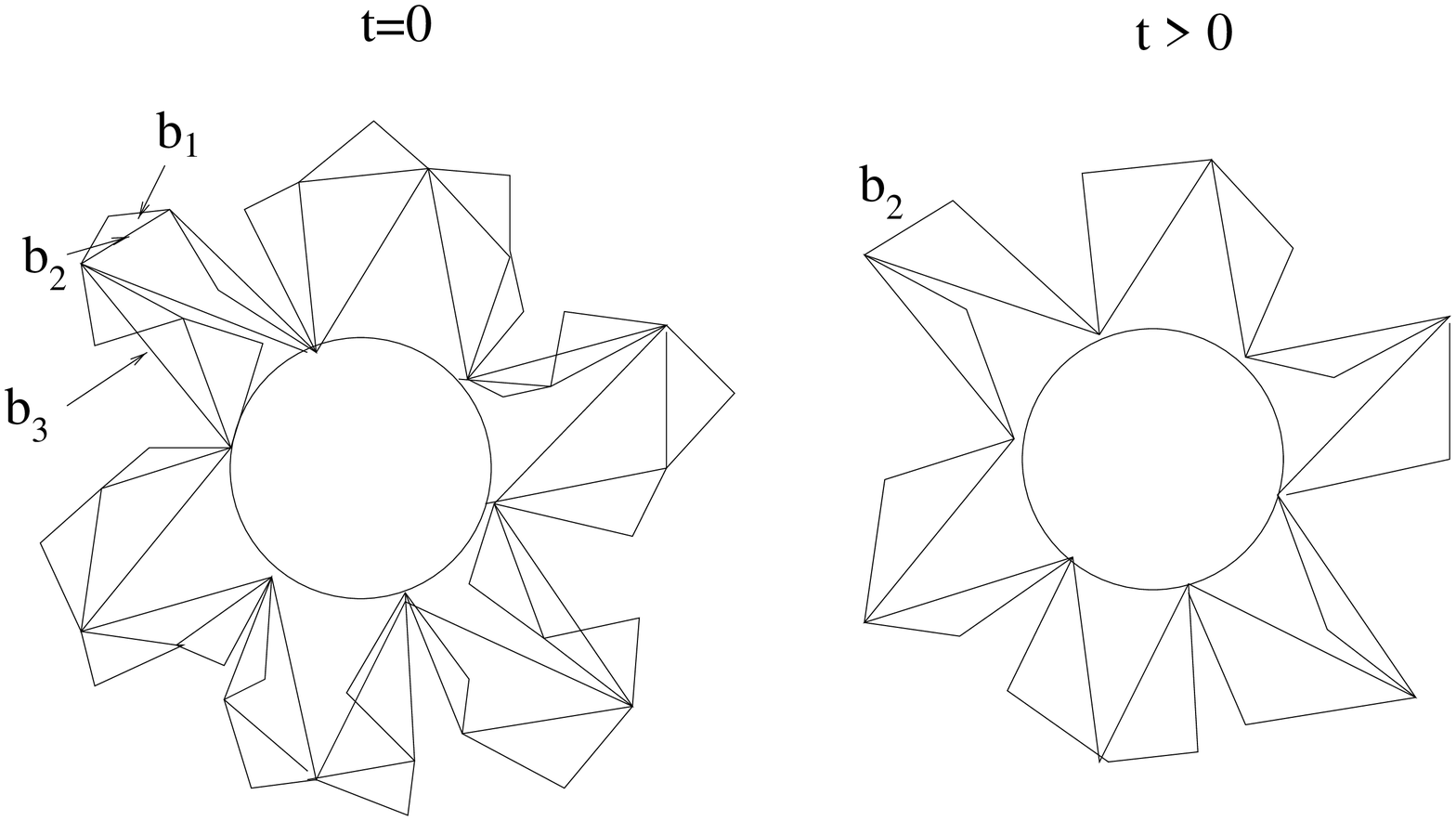}} \vspace*{10pt}
      \caption{\label{fig:gauss}
        Initial configuration of  a chain  can be viewed  as
        hierarchical  fractal in  ring geometry. At time $t$ all
        length scales smaller than a certain $b_2$ becomes collapsed.}
  \end{minipage}
 \end{center}
\end{figure}

To model the collapse process we consider the initial state of globule
formation to be composed by a Gaussian chain of $N$ monomers of size $b_1$. It
can be considered as a fractal. In the states at later time the corresponding
structure is assumed to be the same, generated by a  hierarchical random
process, as illustrated in Fig.\ref{fig:gauss}. This self similar hierarchical
process defines static and dynamic properties by a set of exponents. In each
scale and at each stage, the chain can be again re-expressed by random walks of
$N/g$ coarse grained monomers of size $b(g)$ which contain $g$ original
monomers \cite{abrams01}.  The main effect of the quench to poor solvent
conditions can be viewed as if the Gaussian ($\theta$ -) chain becomes
instantly elastic and collapses in order to minimize its contacts with the
solvent.
At larger times $t$, the Gaussian fractal structure becomes less complex, since
at smaller scales the structure ($b_1$ in Fig.~\ref{fig:gauss}) becomes
collapsed. Then all monomers on length scales smaller than, e.g.  
$b_1$, condense to a scale of length $b_2$.  On the other hand the
overall structure on larger scales remains a random walk of  the  new  coarse grained
monomers of size $b_2$.  After further collapse of $g \sim b_2^2$ monomers,
the longest dimension of each collapsed segment remains as $b_3$.  The
vectorial sum of the net force acting on the each segment from outside of the
segment is zero if the segment belongs to the fractal structure.  In the
linear chain dynamics, the coarse graining picture based on the fractal
structure moves to the  late stage  when the intermediate chain segment
can no longer be considered as a part of the fractal due to the continuous
condensation.  The typical late stage configuration for a linear chain is the
collinear structure where terminal parts of chain experience a net
force\cite{abrams01}.

Due to the absence of end effects in the chosen
ring geometry, the collapse dynamics
is a continuous coarse graining process until the polymer reaches its
compact globule conformation.  We will discuss now  an exponent $x$ for the characteristic time of
the collapse, i.e.  $t^*_{\rm collapse} \sim N^{x}$.  When the chain is
quenched into poor solvent conditions, the initial random walk configuration
contracts immediately.  The energy per each segment of size $R_1$ is $E(R_1)
\sim \tau g k_{\rm B}T \sim \tau (R_1/b)^2 k_{\rm B}T$ where $\tau \sim v_{\rm
  i} - v_{\rm f}$ is the depth of quench ($v_{\rm i}$ and $v_{\rm f}$ are the
initial and the final second virial coefficients correspondingly).  The
contraction results in a finite net force, which can be estimated for a given
segment of length $R_1(g)\sim b\sqrt{g}$ to be
\begin{equation}
f(g)= \frac{d E}{d R_1(g)}  \sim \frac{k_{\rm B}T}{b^2} \tau R_1  \sim
\frac{k_{\rm B}T}{b}\tau \sqrt{g}   .
\label{force}
\end{equation} 
The corresponding velocity of  each segment is  given by
\begin{equation}
u = \frac{f(g)}{\zeta}. 
\label{eq:eq}
\end{equation} 
As we assume the Rouse dynamics, the corresponding friction coefficient of each
segment scales with the number of monomers involved, i.e., $\zeta \sim \zeta_0
g$, where $\zeta_0$ is the bare friction coefficient (determined by the white
noise correlation and the fluctuation dissipation theorem).  Since $R_1^2 \sim
b^2 g$,  the characteristic time, $t^* \sim R_1/u$,  for the collapse
of each (coarse grained) segment 
of length $R_1$ reads
\begin{equation}
t^*(g)  = \frac{g}{\tau}t_0.   
\label{each_segment}
\end{equation}  
Here we  define $t_0 \equiv \zeta_0 b^2/k_{\rm B}T = b^2/D$ where $D$ is the
diffusion constant of a solvent particle and is 
related to $\zeta_0$ by Einstein
relation $D\equiv k_{\rm B}T/\zeta_0$.  Since the chain has self similar
structure on all  scales, this  relation holds  until the chain merges to a
single compact globule.  The total time for the collapse is the time required
for the largest length scale with $N$ monomers.
\begin{equation}
t_{\rm collapse}^* \sim  \frac{N}{\tau} t_0.
\label{total}
\end{equation}
If we  use Zimm dynamics, the friction of the segment of size $R_1$ would
no longer be determined by the number of monomers, but by the geometric size
of the chain \cite{Doi}, i.e., $\zeta_{Z} \sim \zeta_0 R_1$.  The
characteristic time for collapse is $t^*_{Z}\sim \sqrt{g} t_0$.  Note that in
Ref.\cite{abrams01}, the characteristic time for collapse is $t\sim g t_0$ with
Zimm dynamics where the infrastructure of a segment is a series of peal
necklaces due to the capillary instability.  The net force of contraction in
the geometry of necklace is independent of the scale.  The contraction
effectively brings monomers which  belong to the string to the globule.  Therefore
$f_{\rm necklace} \sim k_{\rm B}T/b$. We summarize the characteristic time for
collapse in different regimes in table~\ref{table}.
\begin{center} 
\begin{tabular}{|l|l|l|}\hline
     \phantom{t}  & uniform  $f\sim \sqrt{g}$& necklace $f\sim 1$ \\ \hline
 Rouse $\zeta \sim g$&  $t^u_R \sim g$ &   $t^n_R \sim  g^{3/2}$ \\ \hline 
 Zimm $\zeta \sim \sqrt{g}$  & $t^u_Z \sim \sqrt{g}$  &  $t^n_Z \sim g$\\
 \hline 
\end{tabular}
\label{table}
\end{center}

The preceding scenario of the collapse based on the hierarchical
fractal picture have been discussed first in ref.\cite{abrams01}. It
has been shown there that the theoretical predictions are compatible
with the MD - simulation. However we should emphasize that at a larger 
times, when all pearls are merged to a single cluster,  the driving energy is no longer determined by the fractal
regime, $E \sim \tau g
k_BT$, but mainly ruled by  interfacial effects. This has pointed out by de Gennes
\cite{Gennes2} in his sausage model. 
The dynamics in this regime is determined approximately by changes
of the surface, which are described by
\begin{equation}
E_{\rm inter} \sim \gamma L r \, .
\label{interface}
\end{equation}
In
eq.\ref{interface} $\gamma \sim k_BT/\xi^2$ stands for the surface
tension, $L$ is the ``sausage'' length and $r$ is its
radius. 
The sausage is formed by thermal blobs of diameter $\xi \simeq b/\tau$.
In this regime the collapse can be viewed as a minimization of
the interfacial energy $E_{\rm inter}$ under the constant volume $V
\sim Lr^2$. The
characteristic time which corresponds to this regime is given by
\cite{Gennes2}
\begin{equation}
t^{**} \sim \tau N^2 t_0
\label{t**}
\end{equation}
The crossover between the two  above-mentioned  regimes is detrmined by the
match of the two characteristic times, i.e.,
$t^* \approx t^{**}$ or $\tau^2 N \approx {\rm const}$.

This matching is simply understood by taking into account that the globule
behaves liquid like under the constraint of constant volume. The volume
is given by the number of blobs $N/g_{\rm blob}$, where 
$g_{\rm blob} \sim 1/\tau^2$. Thus we may  conclude that
in the crossover range the number of blobs ${\cal N}_{\rm blob}^{(eq)}
\sim N/g_{\rm blob}$ corresponds to the equilibrium value. De Gennes' 
dynamics is associated therefore with the re-packing of the incompressible
``blob fluid''.

The presence of these  two regimes have been shown by the Brownian
dynamics simulation (see Sec.III B and Fig.6 in ref.\cite{Chang}). It
was found that the first regime becomes faster for the larger
quenching $\tau$ (see eq.(\ref{each_segment})) whereas the second one
become slow as the quenching grows (see eq.(\ref{t**})).

\subsection{Swelling}

When the solvent quality is changed suddenly 
from poor to good or $\theta$ -  conditions
the  globule starts to swell. We consider here for
brevity a quench from a  poor solvent to $\theta$ - solvent
conditions and assume that the globule interior always contains enough 
solvent molecules i.e.,  the solvent is not very  poor ($\tau \le 0$) . Then  one can ignore a specific role of the solvent transport and
assure the homogegeneous globule expansion. In this subsection  we restrict ourselves to  a scaling picture
to set up the main time scales
by employing an ``expanding blob'' picture.  Let us therefore assume that 
during the process of expansion the segments on the length of blob size $\xi$
maintain local equilibrium states, whereas on a larger scales the
system is no longer  in equilibrium.This amounts to the assumption that the initial
state of the globule can be described by  free energy which is proportional
to the number of blobs times the thermal energy \cite{Gennes}. We can 
then write the
free energy for the overall quasi - equilibrium regime in the
following form 
\begin{equation}
F_{\rm glob} \propto k_{B}T \left(\frac{N}{g}\right) \quad,
\label{F11}
\end{equation}
where  the number of monomers in the blob, $g$, is a time dependent
function and still has a Gaussian statistics, i.e.
\begin{equation}
g \propto \left(\frac{\xi}{b}\right)^2 \quad.
\label{g}
\end{equation}
We assume as well that the quasi - equilibrium scaling for the overall 
globule size, $R$, is valid and gives 
\begin{equation}
R \propto \xi \left(\frac{N}{g}\right)^{1/3} \quad.
\label{R11}
\end{equation}
The substitution of eqs.(\ref{g}) and (\ref{R11}) in eq. (\ref{F11})
yields
\begin{equation}
F_{\rm glob} \propto k_{B}T \frac{b^6N^3}{R^6} \quad.
\label{F1}
\end{equation}
To take into account the correct long time limiting behavior, when the
chain size becomes proportional to $b^2 N$, the elastic energy term
should be included. This term counterbalances the over-swelling and
stabilizes the system, so that the whole quasi  - equilibrium free
energy reads
\begin{equation}
F \propto k_{B}T \left( \frac{b^6N^3}{R^6} + \frac{R^2}{b^2N}\right)\quad.
\label{F_whole}
\end{equation}
After that the corresponding equation of motion for $R(t)$, i.e.
\begin{equation}
\zeta \frac{d R}{d t} = - \frac{\delta F}{\delta R}   
\label{zeta}
\end{equation}
takes the following form
\begin{equation}
\frac{N}{D} \frac{dR(t)}{dt} =  \frac{b^6 N^3}{R^7} -  \frac{R}{b^2 
  N}\quad.  
\label{Eq_Motion}
\end{equation}
In eq.(\ref{zeta}) $\zeta \sim \zeta_0 N$ is the Rouse friction
coefficient and in eq.(\ref{Eq_Motion}) $D =  k_{B}T/ \zeta_0$ denotes
the monomer diffusion constant. It should be noted, that we do not use the
Rouse diffusion constant throughout the paper, since we wish to keep track 
of the chain length dependences explicitly. 

The
solution of eq.(\ref{Eq_Motion}) is 
\begin{equation}
R^2(t) = \left[R_0^8 e^{-8Dt/b^2N^2} + b^8 N^4 \left( 1 -
        e^{-8Dt/b^2N^2}\right)\right]^{1/4} \quad.
\label{swell}
\end{equation}
In the  asymptotic limit of $t\rightarrow \infty$, we recover the Gaussian 
chain size, $R^2_{\infty} =  b^2 N$.
The characteristic time for swelling is  determined by the condition: 
$R_g^2 \sim t^{1/2}_{\rm swell} \sim b^2 N$.  The characteristic 
time  $t^*_{\rm swell}$ scales with the  chain length as
\begin{equation}
t^*_{\rm swell} \sim N^2.  
\end{equation}

We will see in Sec.V, devoted to the numerical investigation of the
full equation of motion, that because of coupling of different Rouse
modes the driving force of the swelling is mainly determined by the
large length scale (see e.g. Fig.~\ref{fig:omega}).

In this consideration the chain is treated as ``phantom'',
notwithstanding  the effective interaction is taken into account
through the 
virial coefficients segments still can cross each other.This use of pseudo potentials
does not allow to include topological effects discussed recently.
If, however,  the  effects of topological constraints are  dominant, the chain can
be swollen only by reptation through the channels made of neighboring 
segments.  
The swelling  process  of an  entangled globule needs 
more consideration \cite{Nechaev,Rabin}.


\subsection{Effective Hamiltonians and local time scales}

The scaling shows what can be expected at large length scales only. It is at
first sight difficult to estimate the expected time scales on the level of
Rouse type modes. Nevertheless some remarks to the scale dependence on the
relaxation modes can be drawn. In the  more detailed theory, which is presented
below one can  recover the dynamical scales on the very late
relaxation regimes.   The chain size in globular states scales as $R
\simeq b (N/\tau)^{1/d}$ for scales larger than the thermal blob size
$\xi_{\rm T} = b/\tau$.  A general staring point is the Edwards Hamiltonian
$H({{\bf R}(s)})$, where ${\bf R}(s)$ defines the chain position vectors and
$s$ the contour variable. Any dynamic theory can be formulated through 
the Rouse modes 
which are usually defined by a Fourier transform of the position vectors as
${\bf X}(p) = (1/N)\sum_{s} {\bf R}(s)\exp( i sp)$, where $p$  stands for 
 the  Rouse modes with $p = 2\pi n/N$ and  $n = 0, 1, 2, \dots,N - 1$ \cite{Doi}.  In general the effective Hamiltonian
for a collapsed chain can be written in terms of Rouse modes as a
simple quadratic form
\begin{equation}
\label{effective}
H_{\rm eff} = k_{\rm B}T f(\tau) \sum_{p} p^{2a}|{\bf X}(p)|^{2} \quad,
\end{equation}
where $a$ is determined by the static correlation and the function
$f(\tau)$ is 1 for the good and $\theta$ - solvents  but $f(\tau) \sim \tau^{2/d}$
for poor solvent conditions.  In the case of extended
chains it is easy to show  that $2a=1+2\nu$ ( where  $\nu = 3/(d + 2)$)   whereas in the case of compact
chains  $2a=1+2/d$. The Rouse dynamics for these  effective chain
variables ${\bf X}(p)$ is determined by an Rouse equation of the form
\begin{equation}
\zeta_{0} \frac{\partial {\bf X}(p)} {\partial t} + 
\frac{3k_{\rm B}T}{b^{2}} f(\tau) p^{2a} {\bf X}(p) = 0
\end{equation}
 This  effective Rouse equation set the time scales for the latest
 stages of the dynamical evolution. Namely, for the swelling the
 relaxation spectrum has the form 
 \begin{equation}
\tau_{\rm rel}(p)  \simeq \frac{\zeta_{0}}{p^{1+2\nu}}\quad,
\label{tau1}
\end{equation}
whereas for the collapse it is
\begin{equation}
\tau_{\rm rel}(p) \simeq \frac{\zeta_{0}}{\tau^{2/d}p^{1+2/d}}\quad.
\label{tau2}
\end{equation}
This corresponds naturally to the characteristic  Rouse relaxation
times as we will discuss below in more refined theories
it in Sec. IVB.

\section{Equation of motions for correlators}

\subsection{Model}

In this section we provide a more general formulation of the Langevin dynamics
for a polymer chain  immersed in  the  solvent.  The chain conformation is
characterized by the $d$ - dimensional vector - function ${\bf R}(s,t)$ of the
time $t$ and $s$ ($1 \le s \le N$), the segment position along the chain
contour. The corresponding intra chain Hamiltonian has the following form
\begin{eqnarray}
H=\frac{1}{2}\epsilon\sum_{s=1}^{N}\left[\nabla_{s}{\bf
    R}(s,t)\right]^{2} +  H_{\rm int}\left\{{\bf R}(s,t)\right\} 
\quad,
\label{Hamilton}
\end{eqnarray}
where $\epsilon=dk_B T/b^{2}$ is the elastic modulus with the Kuhn segment length
$b$, $N$ is the general number of segments and the finite difference
$\nabla_{s}{\bf R}_{j}(s,t) = {\bf R}_{j}(s + 1,t) - {\bf R}_{j}(s,t)$ and the
intra-chain interaction Hamiltonian has the form
\begin{eqnarray}  
 H_{\rm int}\left\{{\bf R}(s,t)\right\} &=&\frac{1}{2}
 \sum_{s=1}^{N}\sum_{s'=1}^{N}v({\bf R}(s,t) - {\bf R}(s',t))\nonumber\\
&+&\frac{1}{3!} \sum_{s=1}^{N}\sum_{s'=1}^{N}\sum_{s''=1}^{N}w({\bf
   R}(s,t) - {\bf R}(s',t); {\bf R}(s',t) - {\bf R}(s'',t)) + \dots
\label{Interaction}
\end{eqnarray}
In eq.(\ref{Interaction}) $v({\bf r})$ and $w({\bf r}_1,{\bf r}_2)$ are the
second and the third virial coefficients correspondingly. Let us treat
the   low -  molecular solvent molecules as a separate  component and specify their 
local positions by the vector - function ${\bf
  r}^{(p)}(t)$, where $p = 1, 2, \dots M$ numerates  the number of the solvent
molecules. We denote by $V_{\rm ps}({\bf r})$ and $V_{\rm ss}({\bf r})$ the
polymer - solvent and solvent - solvent interaction potential correspondingly.
After that the whole polymer - solvent dynamics is described by the following
Langevin equations:
\begin{eqnarray}
\zeta_{0}\frac{\partial}{\partial
t}R_{j}(s,t)&-&\epsilon\Delta_{s}R_{j}(s,t) +\frac{\delta}{\delta R_{j}(s,t)} H_{\rm int}\left\{{\bf R}(s,t)\right\} \nonumber\\
&+&{\frac{\delta}{\delta R_{j}(s,t)}\sum_{p=1}^{M} V_{\rm ps}\left({\bf R}(s,t)-{\bf
r}^{(p)}(t)\right)}=f_{j}(s,t)
\label{R}
\end{eqnarray}
and
\begin{eqnarray}
\zeta'_{0}\frac{\partial}{\partial
t}r_{j}^{(p)}(t) &+&\frac{\delta}{\delta
r_j^{(p)}(t)}\sum_{m=1}^M V_{\rm ss}\left({\bf r}^{(p)}(t)-{\bf
r}^{(m)}(t)\right)\nonumber\\&+&\frac{\delta}{\delta
r_j^{(p)}(t)} \sum_{s=1}^{N} V_{\rm ps}\left({\bf r}^{(p)}(t)-{\bf
R}(s,t)\right)={\tilde f}_j^{(p)}(s,t)\quad,
\label{r}
\end{eqnarray}
where $j$ numerates Cartesian components, $\zeta_{0}$ and $\zeta'_{0}$ are
bare friction coefficients of polymer segments and solvent molecules (which
should be of the same order and are put equal in the following) and the second
order finite difference $\Delta_{s}R_{j}(s,t) = R_{j}(s + 1,t) + R_{j}(s -
1,t) - 2R_{j}(s,t)$.

In order to reformulate the Langevin problem (\ref{R}) - (\ref{r}) in a more
convenient form we go to the MSR - functional integral representation
\cite{Rost1}. The generating functional (GF) of the problem has the form
\begin{eqnarray}
Z\left\{\cdots\right\}&=&\int DR_j(s,t)D{\hat
  R}_j(s,t)\nonumber\\
&\times&\exp\left\{\Xi\left[{\bf R}(s,t), \hat
  {\bf R}(s,t)\right] + A_{\rm intra}\left[{\bf R}(s,t),\hat{\bf
  R}(s,t)\right]\right\}\quad,
\label{GF1}
\end{eqnarray} 
where the influence functional (which describes the influence of
solvent molecules on the chain) reads
\begin{eqnarray}
\Xi\left[{\bf R},{\bf \hat
 { R}}\right] &=&\ln\int\prod_{p=1}^{M}D{\bf r}^{(p)}(t)D\hat{\bf
r}^{(p)}(t) \exp\Bigg\{A_{\rm solvent}\left[{\bf r}^{(p)},\hat{\bf r}^{(p)}\right]\nonumber\\
&+&\sum_{p=1}^M\sum_{s=1}^N \int dt
i\hat{R}_{j}(s,t)\frac{\delta}{\delta R_{j}(s,t)}V_{\rm ps}\left({\bf R}(s,t)-{\bf
r}^{(p)}(t)\right)\nonumber\\
&+&\sum_{p=1}^M \sum_{s=1}^N \int dt
i\hat{r}_{j}^{(p)}(t)\frac{\delta}{\delta
  r_{j}^{(p)}(t)}V_{\rm ps}\left({\bf r}^{(p)}(t)-{\bf
R}(s,t)\right)\Bigg\}\quad,
\label{Xi}
\end{eqnarray}
the intra-chain action is given by
\begin{eqnarray}
A_{\rm intra}\left[{\bf R}(s,t),{\hat {\bf R}}(s,t)\right] &=& \sum_{s=1}^N\int dt\Bigg\{i{\hat R}_{j}(s,t)\left[\zeta_{0}\frac{\partial}{\partial t}R_{j}(s,t)-\epsilon\Delta_{s}R_{j}(s,t)\right]\nonumber\\
&+&\frac{\delta}{\delta R_{j}(s,t)} H_{\rm
    int}\left\{R_{j}(s,t)\right\} + k_B T\zeta_{0}\left[i{\hat
      R}_{j}(s,t)\right]^{2}\Bigg\}
\label{testaction}
\end{eqnarray} 
and the solvent action has the form
\begin{eqnarray}
A_{\rm solvent}\left[{\bf r}^{(p)}(t),{\hat {\bf r}}^{(p)}(t)\right]
&=&\sum_{p=1}^{M} \int dt\Bigg\{i{\hat
  r}^{(p)}_{j}(t)\left[\zeta_{0}\frac{\partial}{\partial
    t}r^{(p)}_{j}(t) 
+\sum_{m=1}^{M} \frac{\delta}{\delta r^{(p)}_{j}(t)}V\left[{\bf
    r}^{(p)}(t)-{\bf r}^{(m)}(t)\right]\right]\nonumber\\
&+&k_B T\zeta_{0}\left[i{\hat r}^{(p)}_{j}(t)\right]^{2}\Bigg\}
\label{solventaction}.
\end{eqnarray}
The representation (\ref{GF1}) - (\ref{solventaction}) is a suitable starting
point for further approximations. In particular these expressions are very
convenient for integration over collective solvent variables, which will
leads to the  effective ''actions" for the polymer. We remind that this procedure
is only possible if the solvent particles dynamics is much  faster compared to the
polymer dynamics. While this is in general ensured for long chains, nevertheless this point
has to be treated with a care for the polymer  collapse problem.

\subsection{Self - consistent Hartree approximation}

As mentioned already it is an important to point out how the solvent dynamics
is treated. Indeed we may follow at least two different ways. The solvent can
be considered within the hydrodynamic approximation in the same way  as it was done
in ref. \cite{Fredrick}, where the solvent dynamics was described by an
incompressible Navier - Stokes liquid. This approach leads to a time -
dependent hydrodynamics interaction. The other way is to treat the solvent as
a dynamical background in a random phase approximation (RPA).  Here we restrict
ourselves to the dynamical RPA \cite{Rost1,Rost2} which is well known in 
low molecular liquid dynamics \cite{Boon}. It is also  known \cite{Boon} that the RPA is a mean
field type description in which the free - particle behavior is modified by an
effective interaction. The main drawback of this approximation is that it does
not give the proper hydrodynamic behavior. The reason for this lies in the
fact that RPA neglects  collisions which dominate in the hydrodynamic regime. We
will come to this point in a future publication.

In order to accomplish the calculation for the present purpose we make use of
the the transformation to the collective solvent density \cite{Rost1} and
integrate them out. The level of this procedure will determine the level of
approximation also. We relegate all technical details of this
calculation in the  Appendix A.

The Hartree approximation will then take into account all
mean field diagrams. Naturally the mean field description appears poor in good
solvent conditions, but becomes better in the globule state,
since  fluctuations are  less important under the increasing
globule density.

The GF which is determined by eqs.(\ref{GF2}) and (\ref{A}) is still highly
nonlinear with respect to ${\bf R}(s,t)$ and $\hat{\bf R}(s,t)$. In order to
handle the difficulties which are associated to these we use Hartree - type
approximation. In this approximation the real MSR - action is replaced by the
Gaussian one in such a way that all terms which include more than two fields
${\bf R}(s,t)$ and $\hat{\bf R}(s,t)$ ) are written in all possible ways as
products of pairs of ${\bf R}(s,t)$ or/and $\hat{\bf R}(s,t)$ coupled to the
self - consistent averages of the remaining fields. In ref. \cite{Rost3} it
was shown that if the number of field components is large the Hartree
approximation and the next to the saddle point approximation merge and both
become exact. The resulting Hartree action is a Gaussian functional with
coefficients which could be represented in terms of correlation and response
functions. All calculations are straightforward and details can be found in
the Appendix B of ref. \cite{Rehkopf}. The only difference is that here the
second and third virial terms (two last terms in eq.(\ref{A})) explicitly enter
into the equation. After the collection all terms the final GF reads
\begin{eqnarray}
Z\{\cdots \}&=&\int D{\bf R}D{\hat {\bf R}}\exp\Big\{A_{\rm intra}^{(0)}[{\bf R},{\hat {\bf
    R}}]\nonumber\\
&+&\sum_{s=1}^{N}\sum_{s'=1}^{N}\int_{-\infty}^{\infty}dt\int_{-\infty}^{t}dt'\:i{\hat
  R}_{j}(s,t)R_{j}(s',t')\lambda(s,s';t,t')\nonumber\\
&-&\sum_{s=1}^{N}\sum_{s'=1}^{N}\int_{-\infty}^{\infty}dt\int_{-\infty}^{t}dt'\:i{\hat
  R}_{j}(s,t)R_{j}(s',t)\lambda(s,s';t,t')\nonumber\\
&+&\frac{1}{2}\int_{-\infty}^{\infty}dt\int_{-\infty}^{\infty}dt'\:i{\hat
  R}_{j}(s,t)i{\hat R}_{j}(s',t')\chi(s,s';t,t')\Big\}\quad,
\label{GF3}
\end{eqnarray}
where
\begin{eqnarray}
\lambda(s,s';t,t') &=& \frac{1}{d}G(s,s';t,t')\int\frac{d^{d}k}{(2\pi)^{d}}k^{4}|V_{\rm ps}({\bf k})|^{2}F({\bf k};s,s';t,t')S_{00}({\bf k};t,t')\nonumber\\ 
&-&\int\frac{d^{d}k}{(2\pi)^{d}}k^{2}\left[|V_{\rm ps}({\bf k})|^{2}S_{01}({\bf
    k};t,t') - V_{\rm ps}({\bf k})\delta(t-t')\right]F({\bf
  k};s,s';t,t')\nonumber\\
&+&\sum_{s''=1}^{N}\int\frac{d^{d}kd^{d}q }{(2\pi)^{2d}}w({\bf k},{\bf
  q})F({\bf q};s',s'';t,t)F({\bf k};s,s';t,t)\delta(t - t')
\label{Lambda}
\end{eqnarray}
and
\begin{eqnarray}
\chi(s,s';t,t')=\int\frac{d^{d}k}{(2\pi)^{d}}k^{2}|V({\bf
  k})|^{2}F({\bf k};s,s';t,t')S_{00}({\bf k};t,t')
\label{Chi}.
\end{eqnarray}
In eqs.(\ref{GF3}) - (\ref{Chi}) the response function 
\begin{eqnarray}
G(s,s';t,t')=\left<i{\hat {\bf R}}(s',t'){\bf R}(s,t)\right>
\label{G}
\end{eqnarray}
and the chain  density correlator
\begin{eqnarray}
F({\bf k};s,s';t,t')=\exp\left\{-\frac{k^{2}}{d}Q(s,s';t,t')\right\}
\label{F}
\end{eqnarray}
with 
\begin{eqnarray}
Q(s,s';t,t')\equiv\left<{\bf R}(s,t){\bf R}(s,t)\right>-\left<{\bf
    R}(s,t){\bf R}(s',t')\right> \quad.
\label{Q}
\end{eqnarray}
In eqs.(\ref{GF3}) - (\ref{Chi}) $S_{00}({\bf k};t,t')$ and
$S_{01}({\bf k};t,t')$ are the solvent RPA - density correlation and
response functions correspondingly (see eqs.(\ref{RPA1}) and
(\ref{RPA2}) in the Appendix A). They embody information on the
solvent dynamics.

The pointed brackets denote the self - consistent averaging with the
Hartree - type GF, eq.(\ref{GF3}). Below we will also concern
transient time regimes, so that keeping both time arguments for
correlator $F$ in eq.(\ref{Lambda}) equal to each other does not
necessarily mean that this is a static correlator $F_{\rm st}$. On the 
other hand we assume that the fluctuation - dissipation theorem (FDT)
hold for both chain and solvent correlators, then
\begin{eqnarray}
G(s,s';t-t') = (k_B T)^{-1}\frac{\partial}{\partial t'}Q(s,s';t-t')\:\:\:\:{\rm at}\:\:\: t>t'
\label{FDT1}
\end{eqnarray}
\begin{eqnarray}
S_{01}({\bf k};t-t') = (k_B T)^{-1}\frac{\partial}{\partial t'}S_{00}({\bf
  k};t-t')\:\:\:\:{\rm at}\:\:\: t>t'
\label{FDT2}.
\end{eqnarray}
Note that in eq.(\ref{FDT2})
the units of the correlation function $S_{00}$  and the response function $S_{01}$ are 
different (see the corresponding Appendix A for the notation).

Now we can use eqs.(\ref{FDT1}) and (\ref{FDT2}) in eqs.(\ref{GF3}) -
(\ref{Q}). After integration by parts with respect to time argument 
$t'$, we obtain
\begin{eqnarray}
Z\{\cdots \}&=&\int D{\bf R}D{\hat {\bf R}}\exp\Big\{\sum_{s,s'=1}^{N}\int_{-\infty}^{\infty}dt\int_{-\infty}^{t}dt'\:i{\hat
  R}_{j}(s,t)\left[ \zeta_0 \delta(t - t') + \theta(t -
  t')\Gamma(s,s';t,t')\right]\frac{\partial}{\partial t}R_{j}(s',t')\nonumber\\
&-&\sum_{s,s'=1}^{N}\int_{-\infty}^{\infty}dt\int_{-\infty}^{t}dt'\:i{\hat
  R}_{j}(s,t)\:\Omega(s,s';t)\:R_{j}(s',t)\nonumber\\
&+& k_B T \sum_{s,s'=1}^{N}\int_{-\infty}^{\infty}dt\int_{-\infty}^{\infty}dt'\:i{\hat
  R}_{j}(s,t)\left[ \zeta_0 \delta(t - t') + \theta(t -
  t')\Gamma(s,s';t,t')\right] i{\hat R}_{j}(s',t')\Big\}\quad,
\label{GF4}
\end{eqnarray}
where the memory function 
\begin{eqnarray}
\Gamma(s,s';t)= \frac{1}{k_B T} \int\frac{d^{d}k}{(2\pi)^{d}} \:k^{2}|V_{\rm
  ps}({\bf k})|^{2}F({\bf k};s,s';t)S_{00}({\bf k},t)
\label{Memory}
\end{eqnarray}
and the effective elastic susceptibility
\begin{eqnarray}
\Omega(s,s';t) &=& \epsilon\delta_{\rm ss'}\Delta_{\rm s} -
\int\frac{d^{d}k}{(2\pi)^{d}} \: k^{2}{\cal V}({\bf
  k})\left[F ({\bf k};s,s';t,t) - \delta_{\rm
    ss'}\sum_{s''=1}^{N} F({\bf k};s,s'';t,t)\right]\nonumber\\
&-&\frac{1}{2}\sum_{s''=1}^{N}\int\frac{d^{d}kd^{d}q}{(2\pi)^{2d}} \:
k^2 w({\bf k}, {\bf q})\nonumber\\
&\times&\left[F({\bf k};s,s';t,t)F({\bf q};s'',s';t,t) - \delta_{\rm ss'}\sum_{s'''=1}^{N} F({\bf k};s,s''';t,t)F({\bf q};s''',s'';t,t)\right]\quad.
\label{Omega}
\end{eqnarray}
In eq.(\ref{Omega}) the Fourier components of the effective segment -
segment self - interaction function is given by
\begin{eqnarray}
{\cal V}({\bf k}) = v(k) - \frac{|V_{\rm ps}({\bf k})|^2  \Phi_{\rm
    st}({\bf k})/k_B T}{1 + V_{\rm ss}({\bf k}) \Phi_{\rm st}({\bf
    k})/k_B T}
\quad,
\label{Self-interaction} 
\end{eqnarray}
where $\Phi_{\rm st}({\bf k})$ is the static density correlator for
the free solvent system.
In eq.(\ref{Self-interaction})  the second term results from a coupling with solvent degree of
freedom. The memory function (\ref{Memory}) describes the
renormalization of the Stokes friction coefficient $\zeta_0$ which
originates from a coupling between polymeric and solvent
fluctuations. The effective elastic susceptibility, eq.(\ref{Omega}),
gives an account of the non - dissipative contributions, which arises
not only from the local spring - interaction (the first term in
eq.(\ref{Omega})) but also from effective two and three point
interactions.

\subsection{Equation of motion}

Now we are in a position to derive equation of motion for the time -
displaced correlator
\begin{eqnarray}
C(s,s';t,t') = \left<{\bf R}(s,t){\bf R}(s',t')\right>
\label{TD-correlator}
\end{eqnarray}
as well as for the equal - time correlator
\begin{eqnarray}
P(s,s';t) =  \left<{\bf R}(s,t){\bf R}(s',t)\right> \quad.
\label{ET-correlator}
\end{eqnarray}

 The standard way to derive equations of
motion by starting from the Hartree action (see eq.(\ref{GF3}) is discussed in
Appendix B ref.\cite{Horner}. Using this way for the Hartree action in
eq.(\ref{GF3}) and taking into account FDT  yields
\begin{eqnarray}    
\zeta_0 \frac{\partial}{\partial t}C(s,s';t,t') &-& \sum_{m = 1}^{N} \:
\Omega (s,m;t) C(m,s';t,t')\nonumber\\
 &+&  \sum_{m = 1}^{N}\int_{t'}^{t} \:
\Gamma(s,m;t,\tau) \frac{\partial}{\partial \tau}C(m,s';\tau,t') d \tau = -
2k_B T\zeta_0 G(s',s;t',t) \quad.
\label{EqMotion1}
\end{eqnarray}

In the case of $t' < t$ the r.h.s. of eq.(\ref{EqMotion1}) is zero and 
the time - displaced correlator satisfies the equation
\begin{eqnarray}    
\zeta_0 \frac{\partial}{\partial t}C(s,s';t,t') - \sum_{m = 1}^{N} \:
\Omega (s,m;t) C(m,s';t,t')
 +  \sum_{m = 1}^{N}\int_{t'}^{t} \:
\Gamma(s,m;t,\tau) \frac{\partial}{\partial \tau}C(m,s';\tau,t')d \tau = 0 \quad.
\label{EqMotion2}
\end{eqnarray}
In order to derive the equation for the equal - time correlator
(\ref{ET-correlator}) we recall that
\begin{eqnarray}
\frac{\partial}{\partial t} P(s,s';t) = \left[\frac{\partial}{\partial t}
  C(s,s';t,t')\right]_{t'=t} + \left[ \frac{\partial}{\partial t'} C(s,s';t,t')\right]_{t'=t} 
\label{Recall}
\end{eqnarray}
and the initial condition \cite{Rehkopf}
\begin{eqnarray}
\zeta_0 G(s,s';t+0^+,t') = - d\delta(s - s') \quad.
\label{Initial}
\end{eqnarray}
Let us make the permutation of time moments, $t {\to\atop \gets} t'$,
in eq. (\ref{EqMotion1}). Combining this equation with the original
one and using eqs. (\ref{Recall}) - (\ref{Initial}) in the limit $t =
t' + \epsilon$ at $\epsilon \to 0$, one can derive the result: 
\begin{eqnarray}
\frac{1}{2}\zeta_0 \frac{\partial}{\partial t} P(s,s';t) - \sum_{m = 1}^{N} \:
\Omega (s,m;t) P(s,s';t) = d k_B T \delta(s - s')
\label{EqMotion3}
\end{eqnarray}
It is of interest that in eq.(\ref{EqMotion3}) the memory term is
dropped  out.

As it was discussed in the Introduction , here we restrict
ourselves to the case where the translational invariance along the chain
backbone holds during the collapse (swelling). Generally speaking, the
presence of the ``pearl necklace'' structure breaks down this invariance, so
that the correlator $ P(s,s';t)$ depends not only on the ``chemical distance''
$|s - s'|$ but also on the position along the chain backbone. Even the
interface might violate this invariance because the chain segments on the
surface and in the bulk experience quite different environment. Nevertheless,
in the case when the pearl formation  is fast compared to the dynamics of the
envelop and there positions along the chain are random  these
differences can be averaged out (by preparing an appropriate ensemble
of pearls realizations)  and the effective invariance
still holds. In this situation the Rouse components are the ``good variables''
and it is worthwhile to make the Rouse transformation in the standard way
\cite{Doi}:

\begin{eqnarray}
{\bf X}(p,t) = \frac{1}{N}\sum_{s = 1}^{N} {\bf R}(s,t)
\exp (is p)
\label{Fourier1}
\end{eqnarray}
and
\begin{eqnarray}
{\bf R}(s,t) = \sum_{p = 0}^{2\pi} {\bf X}(p,t)
\exp (- i s p) \quad,
\label{Fourier2}
\end{eqnarray}
where the Rouse mode $p = 2\pi n /N$ at $n = 0, 1, \dots ,N - 1$ and we have used for simplicity the cyclic boundary
conditions. After that the eq.(\ref{EqMotion3}) reads
\begin{eqnarray}
(2 D)^{-1} \frac{\partial}{\partial t} P(p;t) +  \Omega
(p;t) P(p;t) = d N^{-1}\quad,
\label{EqMotionFourier}
\end{eqnarray}
where $D = k_B T/\zeta_0$ is the bare diffusion coefficient and 
\begin{eqnarray}
\Omega(p ; t) &=& \frac{2 d}{b^2} (1 - \cos p) + \frac{N}{k_B T}\: \int
\frac{d^d k}{(2 \pi)^d} \: k^2 {\cal V}({\bf k}) \left[F({\bf k} ; p
;t,t) -  F({\bf k} ; p=0 ;t,t)\right]\nonumber\\
&+& \frac{N^2}{2k_B T}\:\int \frac{d^d kd^d q}{(2 \pi)^2d} \:  k^2 w({\bf
  k},{\bf q})\Big[F({\bf k} ; p;t,t)F({\bf q} ; p  =
  0;t,t)\nonumber\\
&-&  F({\bf k} ; p=0 ;t,t)F({\bf q} ; p=0 ;t,t)\Big]\quad.
\label{OmegaFourier}
\end{eqnarray}
On the other hand, the Rouse transformation of the chain density
correlator has the form
\begin{eqnarray}
F({\bf k} ; p ;t,t) = \frac{1}{N} \sum_{n = 1}^{N} \cos (p
n)\exp\left\{ - \frac{k^2}{d} Q(n ; t,t)\right\}\quad.
\label{F-correlator}
\end{eqnarray}
For the short range segment - segment interaction one can neglect the
wave  vector dependence in ${\cal V}$ and $w({\bf k},{\bf q})$ by
putting ${\cal V} \approx v$ and $w({\bf k},{\bf q}) \approx
w$. Using eq.(\ref{F-correlator}) in eq.(\ref{OmegaFourier}) and
performing the integration over ${\bf k}$ and ${\bf q}$ yields
\begin{eqnarray}
\Omega(p ; t) &=& \frac{2d}{b^2} (1 - \cos p) - v \frac{d^{\frac{d}{2} 
    + 2}}{2 k_B T
  (4\pi)^{\frac{d}{2}}}\sum_{n = 1}^{N}
\: \frac{1 - \cos(p n)}{\left[Q(n,t)\right]^{\frac{d +
      2}{2}}}\nonumber\\
&-& w \frac{d^{d + 2}}{4k_B T
  (4\pi)^{d}}\sum_{n = 1}^{N}\sum_{m =
  1}^{N - n}\: \frac{1 - \cos(p n)}{\left[Q(n,t)\right]^{\frac{d +
      2}{2}}\left[Q(n,t)\right]^{\frac{d}{2}}}\quad.
\label{OmegaFourier1}
\end{eqnarray}
We stress that the equation of motion (\ref{EqMotionFourier}),
(\ref{OmegaFourier1}) for $P(p ;t)$ is highly nonlinear because 
the correlator $Q(n,t)$ depends on  all other  $P(\kappa ;t)$ 
(which provides also a Rouse mode coupling) as follows
\begin{eqnarray}
Q(n,t) &=&   P(n, n;t) -  P(n, 0;t)\nonumber\\
&=& \sum_{\kappa = 2 \pi/N}^{2 \pi} \: \left[1 - \cos(\kappa n)\right] P(\kappa ;t)
\label{Coupling}       
\end{eqnarray}
In equilibrium all equal - time correlators in eqs.(\ref{OmegaFourier1}) 
- (\ref{Coupling}) can be seen as static ones: $P(p ; t) \to
C_{\rm st}(p)$ and $Q(n,t) \to Q_{\rm st}(p)$. Using this limit in
eq.(\ref{EqMotionFourier}) leads to the following  equilibrium equation
\begin{eqnarray}
\left[N C_{\rm st}(p)\right]^{-1} &=& \frac{2}{b^{2}} (1 - \cos p) - v
\frac{d^{\frac{d}{2}+1}}{2 k_B T
  (4\pi)^{\frac{d}{2}}}\sum_{n = 1}^{N}
\: \frac{1 - \cos(p n)}{\left[Q_{\rm st}(n)\right]^{\frac{d +
      2}{2}}}\nonumber\\
&-& w \frac{d^{d+1}}{4 k_B T
  (4\pi)^{d}}\sum_{n = 1}^{N}\sum_{m =
  1}^{N - n}\: \frac{1 - \cos(p n)}{\left[Q_{\rm st}(n)\right]^{\frac{d +
      2}{2}}\left[Q_{\rm st}(n)\right]^{\frac{d}{2}}}\quad.
\label{Static}
\end{eqnarray}

This equation is identical (with the accuracy of prefactors) with the
variational (Euler) equation which we have recently discussed in
ref.\cite{Miglior,Miglior1}. This provides the means for answering one 
of the important questions of whether the Edwards Hamiltonian can be
equally used for dynamical calculations. Within the Hartree
approximation the answer is positive provided that the second and
third virial coefficients are considered as free parameters. In fact,
the Hartree approximation is a dynamical counterpart of the
variational approach \cite{Miglior,Miglior1}.

\section{Early and latest stages of the collapse (swelling)}

Now we are in a position to consider some limiting cases which allow
an analytical investigation. These are obviously very early and latest 
stages of the  transformation. We shall defer the full discussion
which  based on the equation of motion numerical solution until the
next section.

Let us consider the abrupt solvent quality changing between the
initial $v_{\rm i} > 0$ and the final $v_{\rm f} < 0$, which is the case of the
collapse experiment. For the swelling case these are the following:
$v_{\rm i} < 0$ and $v_{\rm f} > 0$.

\subsection{Early stages}

For the very early stage the eq.(\ref{EqMotionFourier}) can be
linearized around the starting  state,  which leads to the following form
\begin{eqnarray}
(2D)^{-1} \frac{\partial}{\partial t} P(p;t) +
\Omega_{\rm st}
(p) P(p;t) = d N^{-1} \quad,
\label{EqMotionLinear}
\end{eqnarray}
where
\begin{eqnarray}
\Omega_{\rm st}(p) &=& \frac{2d}{b^2} (1 - \cos p) - v_{\rm f} \frac{d^{\frac{d}{2} 
    + 2}}{2 k_B T
  (4\pi)^{\frac{d}{2}}}\sum_{n = 1}^{N}
\: \frac{1 - \cos(p n)}{\left[Q_{\rm st}(n)\right]^{\frac{d +
      2}{2}}}\nonumber\\
&-& w \frac{d^{d + 1}}{4 k_B T
  (4\pi)^{d}}\sum_{n = 1}^{N}\sum_{m =
  1}^{N - n}\: \frac{1 - \cos(p n)}{\left[Q_{\rm st}(n)\right]^{\frac{d +
      2}{2}}\left[Q_{\rm st}(n)\right]^{\frac{d}{2}}}\quad.
\label{OmegaFourierStatic}
\end{eqnarray}
In the eq.(\ref{OmegaFourierStatic}) the static correlator $ Q_{\rm
  st}(n)$ should be seen as a coil correlator, i.e. $ Q_{\rm st}(n) =  Q_{\rm
  st}^{coil}(n)$ in the case of collapse and as a globule correlator, 
$Q_{\rm st}(n) =  Q_{\rm st}^{glob}(n)$, in the case of swelling. 

Combining eq.(\ref{OmegaFourierStatic}) with eq.(\ref{Static}), which
is valid in the equilibrium  at $v = v_{\rm i}$, yields
\begin{eqnarray}
\Omega_{\rm st}(p) =  d \left[N C_{\rm st}(p)\right]^{-1} + (v_{\rm i}
- v_{\rm f}) \frac{d^{\frac{d}{2} 
    + 1}}{2 T
  (4\pi)^{\frac{d}{2}}}\sum_{n = 1}^{N}
\: \frac{1 - \cos(p n)}{\left[Q_{\rm st}(n)\right]^{\frac{d +
      2}{2}}} 
\label{OmegaFourierStatic1}
\end{eqnarray}
The very early stage can be characterized by the initial relaxation rate
\begin{eqnarray}
X(p) \equiv  \frac{\partial}{\partial t} 
    P(p;t)\biggr |_{t =0}  = \left[d N^{-1} - \Omega_{\rm st}(p) C_{\rm st}(p)\right](2 D)\quad,
\label{RelaxationRate}
\end{eqnarray}
which with the use of eq.
(\ref{OmegaFourierStatic1}) can be represented in the following form 
\begin{eqnarray}
X(p) = -\frac{(v_{\rm i} -  v_{\rm f})}{\zeta_0}\: \: C_{\rm
  st}(p) \: {\cal F}(p)\quad,
\label{RelaxationRate1}
\end{eqnarray}
where 
\begin{eqnarray}
{\cal F}(p) = \frac{d^{\frac{d}{2} 
    + 2}}{
  (4\pi)^{\frac{d}{2}}}\sum_{n = 1}^{N}
\: \frac{1 - \cos(p n)}{\left[Q_{\rm st}^{(i)}(n)\right]^{\frac{d +
      2}{2}}} \quad.
\label{ConstA}
\end{eqnarray}
The relaxation law for the decrement of the gyration radius  $\Delta R_g^2(t) =
\sum_{p = 2\pi/N}^{2\pi} \left[ P(p;t) -  P(p;0)\right]$
at the early stage takes the form
\begin{eqnarray}
\Delta R_g^2(t) =  -\frac{(v_{\rm i} -  v_{\rm f})}{\zeta_0}\:\:
\Bigl[ \sum_{p = 2\pi/N}^{2\pi}
C_{\rm st}(p) \: {\cal F}(p)\Bigr] \: t \quad.
\label{GerationRad}
\end{eqnarray}
Taking into account the asymptotic behavior,  $\left[C_{\rm
    st}(p)\right]^{-1} \propto p^{1 + 2\nu}$ (for small $p$),
$Q_{\rm st}(n)\propto n^{2 \nu}$ (for large $n$), where
$\nu$ is the Flory exponent, one can obtain for function  ${\cal F}(p)$
the following scaling
\begin{eqnarray}
{\cal F}(p)  \propto p^{\nu(d + 2) - 1}
\label{function_f}
\end{eqnarray}
For the collapse case  $v_{\rm i} > v_{\rm f}$. For quenching from the good
solvent, $\nu =  3/(d + 2)$, and ${\cal F}_c(p)  \propto p^{2}$, whereas for
the quenching from the $\theta$ - solvent $\nu = 1/2$ and  ${\cal F}_c(p)  
\propto p^{d/2}$.

In the case of swelling $v_{\rm i} < v_{\rm f}$. If one heat up the system
starting from the globule state then $\nu = 1/d$ and  ${\cal F}_s(p)  \propto 
p^{2/d}$, whereas for the  $\theta$ - solvent initial state one has
${\cal F}_s(p)  \propto p^{d/2}$.

We will leave the discussion of the relaxation rate $X(p)$ till 
the next section but one can immediately see that $X(p)$ has
the following scaling forms.
For collapse
\begin{equation}
X_c(p)  \propto - \frac{v_{\rm i} - v_{\rm f}}{\zeta_0}
\left\{\begin{array}{l@{\quad,\quad}l}
1/p^{(4-d)/(d+2)} &{\rm good- solvent} \\
1/p^{2 - d/2} &\theta -{\rm solvent} 
\end{array}\right.
\label{Function_R1}
\end{equation}
where the first and the second lines refer to the quenching from the
good and $\theta$ - solvents correspondingly.

For the swelling process we find respectively for different initial starting points
\begin{equation}
X_s(p)  \propto  \frac{v_{\rm f} - v_{\rm i}}{\zeta_0}
\left\{\begin{array}{l@{\quad,\quad}l}
1/p &{\rm poor-solvent} \\
1/p^{2 - d/2} &\theta - {\rm solvent} 
\end{array}\right.
\label{Function_R2}
\end{equation}
Here the first and the second lines are assigned to the heating up from the
poor and  $\theta$ - solvents correspondingly. 

\subsection{Latest stages}

Now the system  is close to the globule state (if the collapse is
under discussion) and we can linearize the equation of motion around
it. In this case eq.(\ref{EqMotionLinear}) is still valid but
$\Omega_{\rm st}(p)$ should be calculated in the globule state, which
comes out of eq.(\ref{OmegaFourierStatic}) after substitution $Q_{\rm
  st}(n) \to Q_{\rm st}^{(gl)}(n)$. In its turn correlator for
the globule state $Q_{\rm st}^{(gl)}(n)$ satisfies the
eq.(\ref{Static}) at $v = v_{\rm f}, C_{\rm st}(p) = C_{\rm st}^{(gl)}(p)$
and $Q_{\rm  st}(n) = Q_{\rm st}^{(gl)}(n)$. After simple
calculations one can obtain
\begin{eqnarray}
\Omega_{\rm st}^{(gl)}(p) =  d \left[N C_{\rm st}^{(gl)}(p)\right]^{-1},
\end{eqnarray}
where $\left[N C_{\rm st}^{(gl)}(p)\right]^{-1}\propto p^{1 + 2/d}$. On
the latest stage only the relaxation mode $p = 2\pi/N$
contributes. Then for $\Delta r_g^2(t) = R_g^2(t) -  R_g^2(\infty)$
we have
\begin{eqnarray}
\Delta r_g^2(t) \propto \exp \left\{- 2 D \:\Omega_{\rm
    st}^{(gl)}\left(p = \frac{2\pi}{N}\right) \:t \right\}\quad,
\end{eqnarray}
i.e. the characteristic relaxation time is given by
\begin{eqnarray}
\tau_{\rm rel} \propto \frac{1}{D} N^{1 + 2/d}\quad,
\label{tau1}
\end{eqnarray}
which agrees with ref. \cite{Pitard1}. Note that we derived this behavior as well 
previously when we had used the effective Hamiltonian for the collapsed chain. Obviously
these crude estimated describe already the main  issues of the dynamic behavior.

The same remark holds for the case of the
large time limit for the swelling in a good solvent ($v_f > 0$). The formal
dynamic theory describes these mode dependence as 
\begin{eqnarray}
\tau_{\rm rel} \propto \frac{1}{D} N^{\frac{d + 8}{d + 2}}\quad,
\label{tau2}
\end{eqnarray}
which is also consistent with ref.\cite{Pitard1}. Again we derived this already
by the effective chain Hamiltonian in Sec. IIC.. 
Physically this appears naturally since the
choice of the exponent in eq.(\ref{effective}) uses implicitly the FDT to determine the
statics correctly. It is interesting
that  eqs.(\ref{tau1}) and  eqs.(\ref{tau2}) can be seen as a
characteristic Rouse time, $\tau_{\rm Rouse}  \sim N^{1 + 2\nu}$ (see
e.g.\cite{PGG}), where the Flory exponent $\nu = 1/d$ or $\nu = 3/(d + 2)$ in the 
case of eqs.(\ref{tau1}) (collapse) and  eqs.(\ref{tau2}) (swelling in a  good solvent)
correspondingly.

\section{Numerical Studies}
\subsection{Collapse}
To analyze these results in more detail we performed numerical studies of the
corresponding equations. We studied so far only the asymptotic behavior, whereas the
intermediate time regimes are not accessible. To do so 
we explicitly computed  the numeric solution  of Eq.~\ref{EqMotionFourier} for chain 
length $N=2^6 -2^{10}$.   For the whole procedure, the three body  
interaction term  is fixed as $v_3 = b^6 \equiv 1$.  
The time scale is measured in units of $t_0 =\zeta_0 b^2/k_B T$.
The monomer density in a globule is determined 
by the balance 
between the two-body attraction and three body repulsion,
$\rho = |v_2|/v_3 = -\tau/b^3$. 
The excluded volume provides the condition for the maximum
density  $\rho_{\rm max}|v_2| \le 1$.

As mentioned earlier, the  chain is phantom, i.e., we use pseudo-potentials 
only which might cause a
faster collapse since the  chain segments are allowed to  pass through each
other.   However, this effect is maybe a  minor correction at least  in the early stages
at the beginning of the
collapse.
At later stages  artifact caused by the phantom assumption
(such as an overshooting on the
relaxation curves) are  expected to be more
pronounced. 
In order to assure the validity of the virial
expansion  and to minimize these artifacts, the solvent quality should
remain in the limit where $v_2^2/v_3 \ll 1$.
Let us now discuss the results in more detail.

\subsubsection {Solvent quality dependence} 

\begin{figure}
\includegraphics[width=10cm]{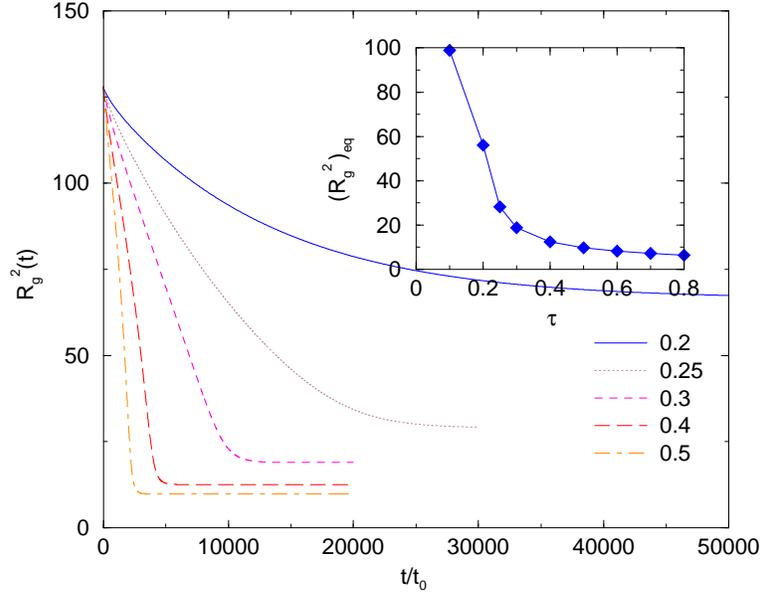}
\caption{\label{fig:sol}
        The mean square  radius of gyration $R_g^2(t)$   as a
        function  of time after quench to 
        poor  solvent conditions from $\theta$ solvent condition 
        where the chain configuration is Gaussian.  
        The equilibrium $R_g^2(t)$ is shown in inset for various solvent
        quality.   The coil to globule
        transition is observed at solvent quality $|\tau|= 0.25$.}
\end{figure}

First we are going to present the dependence on the solvent quality as a first
check of the theory.
The mean squared radius of gyration $R_g^2(t)$  for  various second virial
coefficient  $v_2$ is   plotted  in Fig.~\ref{fig:sol} for $N = 512$. 
Due to the finite size effect the solvent quality which gives 
the coil to globule transition is expected to be  shifted  by
$\sim 1/\sqrt{N}$.
We observe the transition occurs around $\tau \approx -0.25$.   
In  equilibrium, the size of the  globule in the  poor solvent regime  varies 
as $R\sim b(N/|\tau|)^{1/3}$, according to the solvent quality. This is in good
agreement with the scaling predictions.

\subsubsection {Chain length dependence} 

\begin{figure}[h]
\includegraphics[width=10cm]{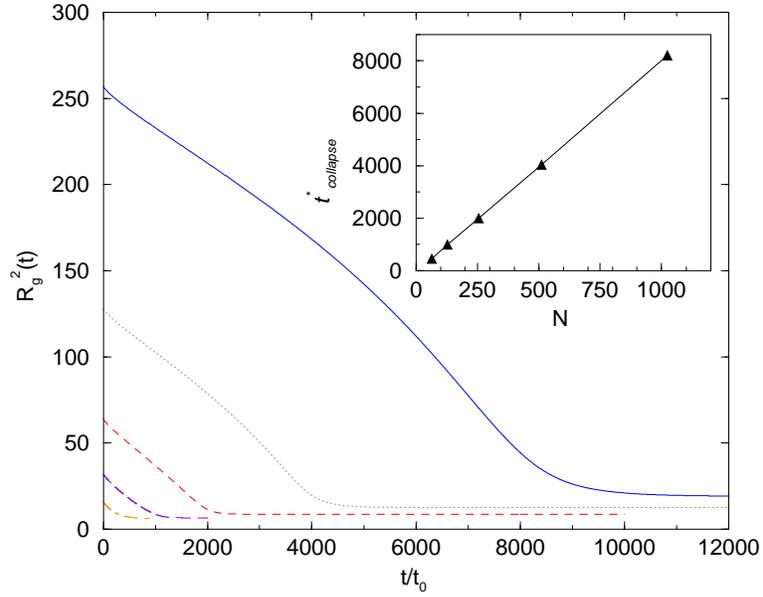}
\caption{\label{fig:chainlength}
        The mean square $R^2_g(t)$ of various chain length $N$  as a
        function of time  after quench to 
        poor  solvent conditions ($\tau = -0.4$) 
        $N=64,128,256,512,1024$ from  bottom to top.
        The inset shows   the total characteristic 
        time for collapse for   each chain   vs. their length  $N$. 
}
\end{figure}

The next issue is the chain length dependence.
In Fig.~\ref{fig:chainlength}, the mean square  radius of gyration
$R^2_g(t)$  during the collapse is shown for various chain size $N$. 
The characteristic 
total collapse time increases linearly 
as chain length $N$, as predicted 
from the  scaling estimate given by eq.(\ref{total}).

\subsubsection{Relaxation times for individual modes} 

\begin{figure}[h]
\includegraphics[width=10cm]{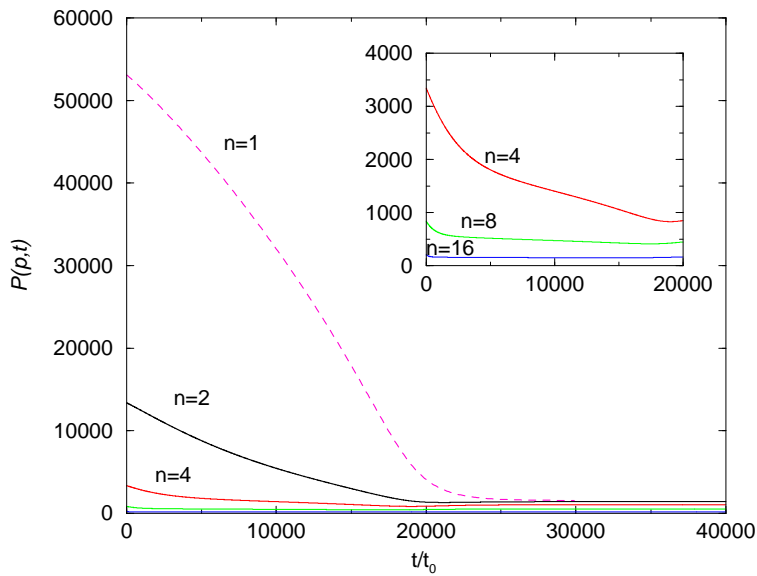} \vspace*{10pt}
      \caption{\label{fig:n1024} {
       The relaxation of each mode $P(p = 2\pi n/N,t)$ during the  collapse transition. 
         }
}
\end{figure}

During the coil - globule transition  the collapse occurs 
in  hierarchical  patterns, such that  the collapsed  segment on  a smaller
length scale can be considered as the unit length of the larger
scale of the chain.  
In the process of the collapse transition  
the active modes disappear one after another. 
We expect such a hierarchical collapse to be visible  in 
the corresponding Fourier mode correlator $P(p,t)$ where $p = 2\pi n/N$ and 
$n = 1, 2, \dots , N-1$.
The characteristic time for mode $p$ 
is related to the relaxation time of the subchain of length  $g= N/n$
After the elapse of  time $t_n$   all length scales less than $g$
are  collapsed (see the Sec.II). 
There exist  $N/g$ collapsed sub-chains at a given time and the hierarchically
larger length structure is a random walk of 
these sub-chains. 
At times less than $t_n$, the contractions  on 
length scale smaller  than $g$  contribute essentially  to the  decrease of 
the correlatator $P(p \sim 1/g,t)$. 
For times $t> t_n$,  further contractions 
from the  larger scales  are  reflected.     
Indeed, the relaxation  behavior of $P(p,t)$  on  smaller length scales (or
larger $n \geq 2$) shows two clearly distinctive regimes, see
Fig.~\ref{fig:n1024}.
However, the longest mode ($n=1$)
shows   only a  single slope until the chain  finds itself in  the globular state.     
We may define the fast decreasing regime as  the first characteristic 
time scale $t_p^*$ for each mode $p$. 
The first characteristic time  $t_p^*$ is  therefore 
related to the  ``internal'' relaxation time of subchain of length  $g= N/n$. 
The subchain relaxation time scales  with its length $g$ as  
$t_p^* \sim g\sim  1/p$ which is consistent with the scaling results
(see Sec.IIA).    
The size of each length scale continue to decrease after time $t_p^*$ 
and this is due to the larger scale contraction.  

It is now instructive  to give a more general estimate for 
the characteristic time $t_p^*$.
According to eq.(\ref{RelaxationRate1}) the  driving force for the
collapse transition scales as $f_p \sim (v_{\rm i} - v_{\rm f}) C_{\rm st}
{\cal F}(p) \sim \tau p^{\nu d - 2} \sim 
\tau /g^{\nu d - 2}$. In  $d=3$ and at $\nu = 1/2$, $f_p \sim
\tau \sqrt{g}$, which agrees with eq.(\ref{force}). In the same manner 
as in Sec.IIA,   
the characteristic time scales as $t_p^* \sim R/u$  where $R$  describes
the chain properties as $R \sim
n^{\nu}$ and $u \sim f_p/\zeta_0 g$, so that  the resulting scaling reads,
\begin{equation}
t_p^* \propto \frac{\zeta_0}{\tau}\; g^{\nu(1 + d) - 1} \quad.
\label{resulting_scale}
\end{equation}
For three dimensions, $d=3$ and $\nu = 1/2$ we recover  the relation, 
$t_p^* \sim \zeta_0 g/\tau$, which is supported by the scaling
analysis in the Sec.IIA.

After the time $t_p^*$, each  Fourier mode decreases slowly (for $n>2$).
The total collapse time for each mode is identical with that of the
first  characteristic time for the longest mode, $t_{p=2\pi/N}^*$.      
The slope  decrease after time $t_p^*$ is controlled (because of modes 
coupling) by the contraction
on  longer scales (or smaller $p$ modes).   
The fast relaxation time  for the  second largest
mode for the chain length $N$ 
coincides with the total collapse time for a chain of the corresponding
half length (see again the Sec. II).   
Naturally, the longest relaxation mode has only one characteristic
regime.  

\subsection{Swelling}

We turn now to the the case of swelling and present the numerical
results of the globule 
expanding  after a quench from poor to a good (or $\theta$ - solvent) condition. 
The initial configuration is now assumed to be a compact globule which is 
prepared by a  poor solvent with a certain negative second virial coefficient. 
Contrary to the collapse dynamics, which based on the hierarchically
crumpled fractal picture, the swelling can be conceived as an
homogeneous extension of the different Rouse modes.
  
\begin{figure}[h]
\includegraphics[ width=10cm]{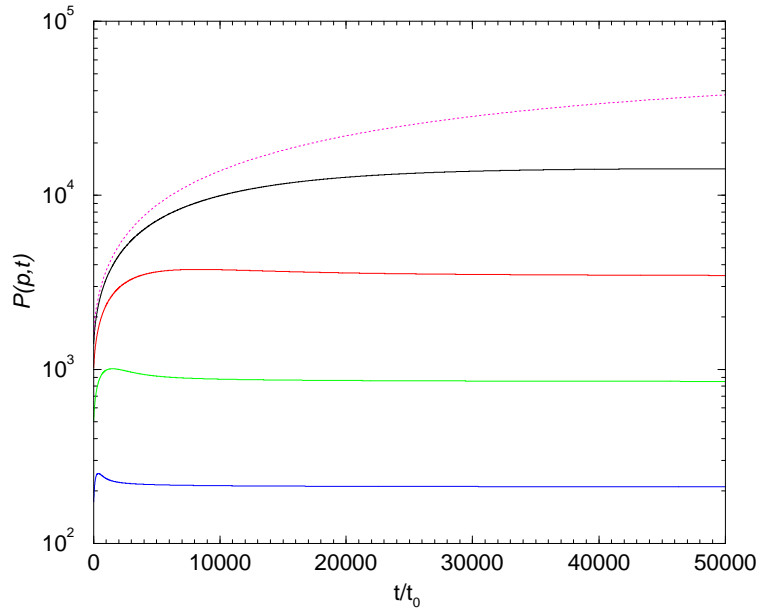}
      \caption{\label{fig:swellp}
      The  mode correlation function $P(p,t)$  for 
      each mode   $pN/2\pi$=1,2,4,8,16 from  top to
      bottom.  The solvent quality is switched from $v_{\rm i} = -0.3b^3$ to 
$v_{\rm f} = -0.1 b^3$. The chain length is N=1024. 
         }
\end{figure}

The time dependence of $ P(p,t)$  is shown in Fig.~\ref{fig:swellp}
for different values of the mode index $p$.  Small length scales (or large $p$)
relax fast, in the sense that they are saturated faster,  while larger 
scales grow  slower. There is no clear 
dynamic exponent observed for the subchain relaxation time in the numerical solution.
In the beginning of the swelling, overshooting is 
observed for  large $p$  values.    
This is an obvious  artifact of a phantom chain.
The monomers move out from the dense globule explosively as soon as 
the solvent condition is switched to a theta solvent.  The phantom chain can 
pass through surrounding  dense phase without being  hindered by the
topological constraint.   
The overshooting shows up when  the  size of the  subchain  of $g$
monomers $r(g)$ is comparable to the  boundary size of the total chain
$R(t)$.   It appears also that the virial expansion is not simply applicable for
at least the early stage of swelling.  

The growth  of the  overall size $R_g^2(t) \sim t^{z}$ is also computed.  
The dynamic  exponent is approximately  $z \sim  1/2$. 
The total relaxation time   $t^*_{swell}$ grows as $N^2$ as the chain length increases. 

\begin{figure}[h]
\includegraphics[ width=10cm]{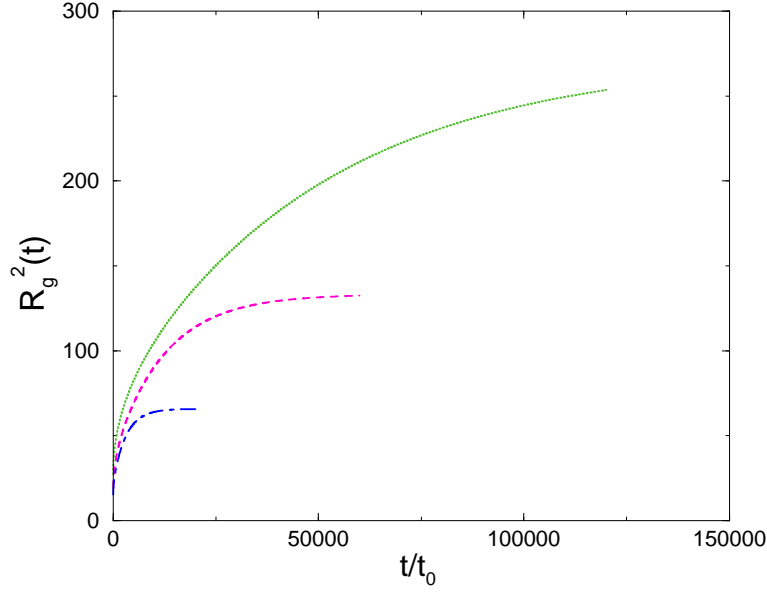} 
      \caption{\label{./figs/fig:swellrg}
      $R_g(t)$ for chain length N = 256, 512, 1024.
         }
\end{figure}

\begin{figure}[h]
\includegraphics[ width=10cm]{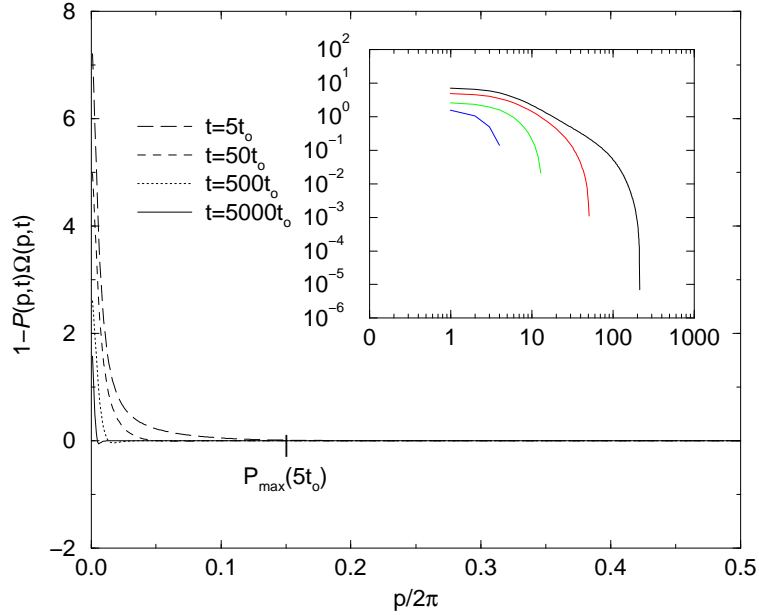} 
      \caption{\label{fig:omega}
      The  relaxation rate  $1-\Omega (p) P(p)/d$ for half cycle of
      $p$ with  chain length N=1024. The contribution to
      the swelling  comes  from  each mode $p < p_c$ at given time.
      The positive  value regimes  are active mode.      
      The inset shows the log-log plot of the relaxation rate. 
         }
\end{figure}

In order to demonstrate  how different modes contribute to the swelling, it is useful to
calculate the normalized relaxation rate $\overline{X}(p,t) = 1 - P (p,t)
\Omega (p,t)/d$ as a function of wave vector $p$. We  have discussed
already (see Sec.IV) the asymptotic forms, i.e. at $p \ll 1$,  of this function  at
the initial time moment. In Fig.~\ref{fig:omega}, we
compute the relaxation rate $\overline{X}(p,t)$ for the swelling condition.
When the system is suddenly quenched to the $\theta$ solvent, the compact
globule conformation is unstable and far from the equilibrium.  The relaxation
rate $X(p,t)$ reflects the discrepancy between the desired conformation and
the conformation at given time $t$.  If the system tends to the  equilibrium, the
relaxation rate obviously vanishes.  The contribution to the total swelling
from different modes $p$ changes as swelling proceeds.  The length scales
smaller then the blob  size do not contribute  in the swelling.  We may define
$p_{\rm max} = 2\pi /g$ as the Rouse index of the  shortest  active mode at
$t=0$, where  $g$ is again the number of chain units in the Gaussian
blob. (The larger $p>
p_{\rm max}$ do not contribute).  As the chain swells, the size of the boundary is
growing, but  on the other hand,  the position of $p_{\rm max}(t)$
decreases accordingly.  In Fig.~\ref{fig:omega} one can see the
shifting  of $p_{\rm max}(t)$ to the smaller value of $p$  in the
course of time. These findings are consistent with the ``expanding
blob'' picture which we discussed in the Sec.II B.

\newpage

\section{Conclusion}

Using the MSR - generating functional technique and the self - consistent
Hartree approximation we have derived the equation of motion for the time
dependent monomer - to - monomer correlation function. The numerical solutions
of this equation for the Rouse modes and the gyration radius $R_g$ in the
cases of collapse and swelling regimes have been discussed. It has been shown
that the simple scaling arguments for the collapse based on the hierarchically
crumpled fractal picture match pretty well with our numerical
findings. The size of the crumples (or the size of the collapsed
segment of the chain) grows  successively, so that the characteristic 
time of the collapse changes  linearly with the length of the
collapsed segment. Swelling,
on the contrary, goes homogeneously, with the  Rouse modes relaxing
with  different relaxation rates. The  spectrum of the relaxation rate
spans the interval between the minimal $p_{\rm min} = 2\pi/N$   and
maximal $p_{\rm max} = 2\pi/g$ Rouse mode indices, where the ``expanding blob'' length,
$g$, is a growing function of time.

In this paper we have restricted ourselves to the ring polymer. The
presence of free ends in the open polymer chain brings some special
features mainly on the later stage of the collapse. This question have 
been discussed in the number of papers \cite{abrams01,ostrovsky}. In
particular in ref. \cite{abrams01} it has been shown that the late
stage configuration consists of two globules with connecting bridge
between them (dumb - bell structure). In this case the dynamics is
determined by the bridge tension between the globules and the
hydrodynamic friction experienced by the globules. 

We should stress that the whole consideration ignores up to now two important
things: (i) hydrodynamical interaction and (ii) topological (or entanglements)
effects. The hydrodynamical interaction can be simply taken into account by
treating the solvent as a incompressible Navier - Stokes liquid coupled with
chains monomers \cite{Fredrick}. Within the self - consistent Hartree
approximation the effective friction coefficient depends from the monomer - to
- monomer correlation function, $Q(n,t)$, so that the closed equation of
motion for the Rouse modes correlator can be solved numerically. This point
will be discussed in detail in a forthcoming publication.

The systematic inclusion of topological constraints in the dynamics of
collapse (or swelling) is a much more complicate theoretical problem. It was
argued in ref.\cite{Nechaev} that at the chain length $N \gg N_{\rm e}$ (where
$N_{\rm e}$ is an effective entanglement length) the topological constraints
become essential. The self similar process of crumpling which we have
discussed in Sec.IIA is restricted also by topological constraints as soon as
a collapsed segment is longer then $N_{\rm e}$. These segments have a fixed
topology and in a sense are in a partially equilibrium state called ``fractal
crumpled globule''. The characteristic time of of this first stage of collapse
is $t_{\rm collapse} \approx t_{\rm collapse}^{(0)} (1 + b^6 /N_{\rm e}w)$ ,
where $w$ is a third virial coefficient and $t_{\rm collapse}^{(0)}$ is the
collapse time without the topological effects. At the subsequent stage the
``crumpled globule'' relaxes to the full equilibrium via the penetration of
the chain ends through the globule and many knots formation. This stage has a
reptational mechanism and as a result the characteristic relaxation time is
scaled as $t_{\rm top} \sim N^3$. Unfortunately, nowadays it is not quite
clear how to incorporate the topological constraints in the equation of
motion, so that this is a challenging problem for the future discussion.
Finally we can mention as an interesting possible application of our
approach the problem of the globule dissolution under an external
force \cite{Kreitmeier}. This problem is also important for the
interpretation of the stress - strain relations in polymer networks
\cite{Cifra}. 
\section*{Acknowledgments}
The authors have benefited from discussions with Albert Johner. V.G.R. and T.A.V.
acknowledge financial support from the Laboratoire Europ\'een Associ\'e (L.E.A.)
N-K.L. and T.A.V.
appreciate  financial support from the German Science Foundation (DFG, 
Schwerpunkt Polyelektrolyte) and support form the Ministery of Research (BMBF) via
the Nanocenter Mainz. 

\begin{appendix}
\section{Integration over solvent variables}
In order to accomplish the RPA -  calculation let us make as usual
\cite{Rost1} the transformation to the collective solvent density
\begin{eqnarray}
\rho({\bf r},t)=\sum_{p=1}^{M}\:\delta\left({\bf r}-{\bf
    r}^{(p)}(t)\right)
\label{rho}
\end{eqnarray}
and response field density
\begin{eqnarray}
\pi({\bf
  r},t)=\sum_{p=1}^{M}\sum_{j=1}^{d}\:i{\hat r}^{(p)}_{j}(t)\nabla_{j}\delta\left({\bf r}-{\bf r}^{(p)}(t)\right)\label{pi}\quad.
\end{eqnarray}
  These transform the influence functional (\ref{Xi}) to the
form
\begin{eqnarray}
\Xi\left[{\bf R},{\hat {\bf R}}\right] &=& \ln \int D\rho({\bf k},t)D\pi({\bf
  k},t)\nonumber\\
&\times&\exp\Bigg\{W[\rho,\pi]-\int dt\int\frac{d^{d}k}{(2\pi)^{d}}\pi (-{\bf
  k},t)\rho({\bf k},t)V_{\rm ss}({\bf k})\nonumber\\
&+& \sum_{s=1}^N\int dt\:i{\hat
  R}_{j}(s,t)\int\frac{d^{d}k}{(2\pi)^{d}}ik_{j}V_{\rm ps}({\bf k})\rho({\bf k},t)\exp\{i{\bf
  k}{\bf R}(s,t)\}\nonumber\\
&-&\sum_{s=1}^N \int dt\int\frac{d^{d}k}{(2\pi)^{d}}\pi ({\bf
  k},t)V_{\rm ps}(-{\bf
  k})\exp\{i{\bf k}{\bf R}(s,t)\}\Bigg\}\dots,
\label{Xirho}
\end{eqnarray} 
where the functional  $W$ depends only on properties of the free system
\begin{eqnarray}
W\{\rho,\pi \}&=&\ln \int\prod_{p=1}^{M}D{\bf r}^{(p)}D{\hat {\bf
    r}}{(p)}\exp\left\{A_{\rm solvent}^{(0)}\left[{\bf r}^{(p)},{\hat {\bf
        r}}^{(p)}\right]\right\}\nonumber\\
&\times&\delta\left(\rho({\bf r},t)-\sum_{p=1}^{M}\:\delta\left({\bf r}-{\bf
    r}^{(p)}(t)\right)\right)\nonumber\\
&\times&\delta\left(\pi ({\bf
  r},t)-\sum_{p=1}^{M}\sum_{j=1}^{d}\:i{\hat
  r}^{(p)}_{j}(t)\nabla_{j}\delta\left({\bf r}-{\bf
    r}^{(p)}(t)\right)\right) \ .
\label{W}
\end{eqnarray}
Here $A_{\rm solvent}^{(0)}$ is the free solvent action. Following ref.
\cite{Rost1,Rost2} one can expand $W\{\rho,\pi \}$ up to the second order with
respect $\rho$ and $\pi$, which formally corresponds to the dynamical RPA.
Then  the solvent variables in (\ref{Xirho}) can be integrated over and
for GF we obtain the following result
\begin{eqnarray}
Z\left\{\cdots\right\}=\int DR_j(s,t)D{\hat
  R}_j(s,t)\: \exp\left\{ A_{\rm eff}\left[{\bf R}(s,t),\hat{\bf
  R}(s,t)\right]\right\}\quad,
\label{GF2}
\end{eqnarray}
where 
\begin{eqnarray}
A_{\rm eff}\left[{\bf R},{\hat {\bf R}}\right]&=&A_{\rm intra}^{(0)}\left[{\bf R},{\hat {\bf
        R}}\right]\nonumber\\
&+&\frac{1}{2}\int_{1}\int_{1'}\:i{\hat
  R}_{j}(1)\int\frac{d^{d}k}{(2\pi)^{d}}k_{j}k_{p}|V_{\rm ps}({\bf
  k})|^{2}\exp\left\{i{\bf k}\left[{\bf R}(1)-{\bf R}(1')\right]\right\}i{\hat
  R}_{p}(1')S_{00}({\bf k},t-t')\nonumber\\
&-&\int_{1}\int_{1'}\:i{\hat
  R}_{j}(1)\int\frac{d^{d}k}{(2\pi)^{d}}ik_{j}|V_{\rm ps}({\bf
  k})|^{2}\exp\left\{i{\bf k}\left[{\bf R}(1)-{\bf
      R}(1')\right]\right\}S_{01}({\bf k},t-t')\nonumber\\
&+&\int_{1}\int_{1'}\:i{\hat
  R}_{j}(1)\int\frac{d^{d}k}{(2\pi)^{d}}ik_{j}v({\bf k})\exp\left\{i{\bf k}\left[{\bf R}(1)-{\bf
      R}(1')\right]\right\}\delta(t-t')\nonumber\\
&+&\frac{1}{2}\int_{1}\int_{1'}\int_{1''}\:{\hat
  R}_{j}(1)\int\frac{d^{d}kd^{d}q }{(2\pi)^{2d}}ik_{j}w({\bf k},{\bf
  q})\nonumber\\
&\times&\exp\left\{i{\bf k}\left[{\bf R}(1)-{\bf
      R}(1')\right] + i{\bf q} \left[{\bf R}(1')-{\bf
      R}(1'')\right] \right\}  \delta(t-t') \delta(t'-t'')\quad,
\label{A}
\end{eqnarray}
where we have used a short hand notations: $\int_{1} \equiv
\sum_{s=1}^{N}\int_{-\infty}^{\infty}\:dt$, $1 \equiv (s,t)$ and
$A_{\rm intra}^{(0)}$ is the free chain action. In eq.(\ref{A})
$)S_{00}({\bf k},t)$ and $S_{01}({\bf k},t)$ stands for the solvent
RPA - correlation and response functions correspondingly. They take
especially simple forms after Fourier transformation with respect to
time argument. This can be written as
\begin{eqnarray}
S_{00}({\bf k},\omega) &=& \frac{\Phi_{00}({\bf k},\omega)}{\left[ 1 +
    V_{\rm ss}({\bf k}) \Phi_{10}({\bf k},\omega)\right]\left[ 1 +
    V_{\rm ss}({\bf k}) \Phi_{01}({\bf k},\omega)\right]} 
\label{RPA1}
\end{eqnarray}
and
\begin{eqnarray}
S_{01}({\bf k},\omega) &=& \frac{\Phi_{01}({\bf k},\omega)}{\left[ 1 +
    V_{\rm ss}({\bf k}) \Phi_{01}({\bf k},\omega)\right]} ,
\label{RPA2}
\end{eqnarray}
where $\Phi_{00}({\bf k},\omega),\Phi_{01}({\bf k},\omega) = \Phi_{01}^{*}({\bf
  k},\omega)$ are correlation and response function for the free
solvent correspondingly. 

\end{appendix}

\end{document}